\documentclass[twocolumn,aps,prl,amsmath,superscriptaddress,notitlepage,longbibliography]{revtex4-1}
\usepackage{amssymb}
\usepackage{mathrsfs}
\usepackage{graphicx}
\usepackage{float}
\usepackage[caption=false]{subfig}
\usepackage{scalerel}
\usepackage{epstopdf}
\usepackage[normalem]{ulem}
\usepackage{verbatim}
\usepackage{xcolor}
\usepackage{bm}
\usepackage{dcolumn}% Align table columns on decimal point
\usepackage{threeparttable}

\usepackage[utf8x]{inputenc}
\usepackage{braket}
\usepackage{orcidlink}

\usepackage{algorithm}
\usepackage{algpseudocode}
\usepackage{amsmath,epsfig,amssymb,bm,dsfont}

\usepackage{listings}
\begin{document}

\newcommand{\norm}[1]{\left\lVert#1\right\rVert}
\newcommand{\ad}[1]{\text{ad}_{S_{#1}(t)}}

\title{A fast and exact approach for stabilizer R\'enyi entropy via the XOR–FWHT algorithm}

\author{Xuyang Huang}
\thanks{These two authors contributed equally to this work.}
\affiliation{Institute for Quantum Science and Technology, Shanghai University, Shanghai 200444, China}

\author{Han-Ze Li}
\thanks{These two authors contributed equally to this work.}
\affiliation{Institute for Quantum Science and Technology, Shanghai University, Shanghai 200444, China}
\affiliation{Department of Physics, National University of Singapore, Singapore 117542, Singapore}

\author{Ching Hua Lee}
\email{phylch@nus.edu.sg}
\affiliation{Department of Physics, National University of Singapore, Singapore 117542, Singapore}

\author{Jian-Xin Zhong}
\email{jxzhong@shu.edu.cn}
\affiliation{Institute for Quantum Science and Technology, Shanghai University, Shanghai 200444, China}

\date{\today}

\begin{abstract}
Quantum advantage is widely understood to rely on key quantum resources beyond entanglement, among which nonstabilizerness (quantum ``magic'') plays a central role in enabling universal quantum computation.
However, the exact evaluation of the second-order stabilizer R\'enyi entropy for generic many-body quantum states remains computationally challenging, with brute-force methods scaling as $\mathcal O(8^N)$ for an $N$-qubit state.
Here we develop a deterministic and exact algorithm that reduces this cost to $\mathcal{O}(N4^N)$ while retaining natural parallelism.
This advance enables high-precision exact calculations for generic state vectors at medium system sizes, and provides a practical tool for investigating the scaling, phase structure, and nonequilibrium dynamics of quantum magic in many-body systems.
\end{abstract}

\maketitle

\noindent{\textcolor{purple}{\textit{Introduction.---}}} Quantum advantage is commonly understood within the framework of resource theories, in which the key resources include quantum entanglement~\cite{Entanglement1,Entanglement2} and \emph{nonstabilizerness}~\cite{RevModPhys.91.025001,Veitch_2014,PhysRevLett.128.050402,PRXQuantum.3.020333}, the latter also known as quantum \emph{magic}. Entanglement alone, however, is not sufficient to guarantee quantum advantage, since even highly entangled stabilizer states remain efficiently simulable on a classical computer under the Gottesman--Knill theorem~\cite{gottesman1998heisenbergrepresentationquantumcomputers, GT}. Magic is therefore regarded as an independent resource necessary for quantum advantage: it characterizes how far a quantum state is from the class of stabilizer states and is associated with non-Clifford resources, which are costly in experiments and in fault-tolerant architectures.~\cite{Preskill2018NISQ,PRXQuantum.3.020333,Quantinuum2025,Aasen2025,Peham2025}

In the recent years, nonstabilizerness in many-body quantum systems has attracted growing interest. At the interface of quantum information and nonequilibrium many-body physics, issues such as scrambling, operator growth, and complex quantum dynamics have been extensively explored~\cite{HaydenPreskill2007JHEP,SekinoSusskind2008JHEP,HosurQiRobertsYoshida2016JHEP,vonKeyserlingkRakovszkyPollmannSondhi2018PRX,CH2,KhemaniVishwanathHuse2018PRX,Swingle2018NatPhys,CH1,HuangLiHuseChan2023Scholarpedia}. In this context, magic and related quantifiers of nonstabilizerness have been used to characterize many-body quantum states and their dynamical behavior~\cite{PRXQuantum.3.020333,Tarabunga2024criticalbehaviorsof,PRXQuantum.4.040317,PhysRevB.111.L081102,LiuZhangYinZhang2024PRL,LiuZhangYinZhangYao2024arXiv,YuLiZhang2025CPL,PhysRevResearch.7.L012011,Chen2025subsystem,PhysRevB.110.064323,PhysRevLett.132.240402,advs.202513868}. A broad range of magic measures has been proposed, including quantifiers based on quasiprobability representations, robustness-type measures, and measures related to operational or simulation overhead~\cite{Veitch_2014, WOOTTERS19871, Veitch_2012, PhysRevLett116250501, Bravyi, PhysRevLett.124.090505, Heimendahl2021stabilizerextentis}. However, many of these measures typically entail \emph{exponential} complexity in numerical evaluation and are therefore only applicable to small-scale systems. By contrast, the stabilizer R\'enyi entropies (SREs) provide a complementary framework for quantifying nonstabilizer resources~\cite{PhysRevLett.128.050402,PhysRevA.107.022429,PhysRevA.110.L040403}. Among them, the second-order SRE (2-SRE) is the simplest nontrivial case, possessing an explicit quartic structure in its representation in terms of Pauli expectation values. This quantity has been widely studied in a variety of settings, including quantum phase transitions~\cite{PhysRevB.107.035148, PRXQuantum.4.040317, 10.21468/SciPostPhys.15.4.131, PhysRevA.110.022436, PhysRevB.111.L081102, pyzr-jmvw, 1tyr-rlbb} and nonequilibrium dynamics~\cite{maity2025local, bejan2025magic, CH3,mello2025magic, PhysRevA.108.042407,myrb-nyhf, PhysRevB.111.054301, 1jzy-sk9r, y9r6-dx7p, Haug2025probingquantum,d7tm-9hkp, jplh-zl35, hou2025stabilizer}.

Direct numerical evaluation of the 2-SRE remains highly expensive. The summation runs over the full set of Pauli strings, whose cardinality grows exponentially with system size as $|\mathcal P_N|\sim 4^N$; for a generic state-vector input, the brute-force approach has overall complexity $\mathcal O(8^N)$, so that exact calculations for spin-$\frac12$ systems are in practice usually restricted to relatively small system sizes, typically $N\lesssim 14$~\cite{njgn-fksh,10.21468/SciPostPhys.19.6.159,PhysRevC.111.034317,trigueros2025nonstabilizerness}.
To overcome this bottleneck, several computational strategies have been developed for the 2-SRE and related quantities. Tensor-network approaches are effective when the quantum state admits a low-entanglement representation or a controlled bond dimension, although their accuracy is limited by the quality of the state representation and truncation errors~\cite{Haug2023stabilizerentropies, PhysRevB.107.035148, PhysRevLett.131.180401, PhysRevLett.133.010601, PhysRevLett.133.010602,tarabunga2025efficientmutualmagicmagic,PRXQuantum.5.030313,liu2025stabilizerrenyientropytranslationinvariant,PRXQuantum.6.010345,hoshino2025stabilizerrenyientropyconformal,hoshino2025stabilizerrenyientropyencodes,PhysRevB.109.174207}. Sampling and Monte Carlo methods avoid explicit enumeration of all $4^N$ terms, but at the price of statistical uncertainty~\cite{PhysRevB.111.085144,PRXQuantum.4.040317, Tarabunga2024magicingeneralized,crew2025learningmagicschwingermodel,sinibaldi2025nonstabilizernessneuralquantumstates}. For states with special algebraic structure, such as Gaussian states, substantially faster evaluations can be achieved, but the applicability is then restricted to these structured families~\cite{PhysRevA.108.042407, collura2025nonstabilizernessfermionicgaussianstates,mx8t-l4hf,s1, PhysRevA.106.042426,rajabpour2025stabilizershannonrenyiequivalenceexact,wang2025magictransitionmonitoredfree}.
Overall, existing methods generally lack full universality, as they are often constrained by entanglement structure, the form of the state representation, or numerical accuracy. It is therefore highly desirable to develop a numerical method for computing the 2-SRE that does not rely on \emph{a priori} structural assumptions about the quantum state, thereby retaining broad applicability while simultaneously achieving high accuracy and computational efficiency.

In this letter, we fill this gap by presenting a deterministic and exact algorithm for evaluating the 2-SRE of a generic $N$-qubit pure state represented as a state vector.
We systematically rewrite the Pauli-string summation in the 2-SRE expression into a bitstring form and reveal convolution and correlation structures over the group $\mathbb Z_2^N$, namely XOR-convolutions~\cite{Ueno_Ito_Todo_Inoue_Minematsu_Ishikawa_Homma_2025} and correlations over bitstrings, which enables us to exploit the orthogonality identities and the convolution theorem associated with the Walsh-Hadamard transform.
This yields an exact formula amenable to efficient numerical implementation, in which the computation reduces to repeated fast Walsh-Hadamard transforms (FWHTs), pointwise operations, and index reshuffling.
Accordingly, we refer to the algorithm proposed in this work as the {XOR-FWHT} algorithm.
The algorithm requires repeated FWHTs of length $d$, where $d=2^N$ is the Hilbert-space dimension, achieves a runtime scaling of $\mathcal O(N4^N)$ with modest additional memory overhead, and is naturally parallelizable over the shift index.
We show that this reduction yields a substantial computational speedup over brute-force Pauli enumeration and enables exact evaluation of the 2-SRE up to $N\!\sim\!20$ for generic state vectors.
\\

\begin{figure}[bt]
\includegraphics[width=\linewidth]{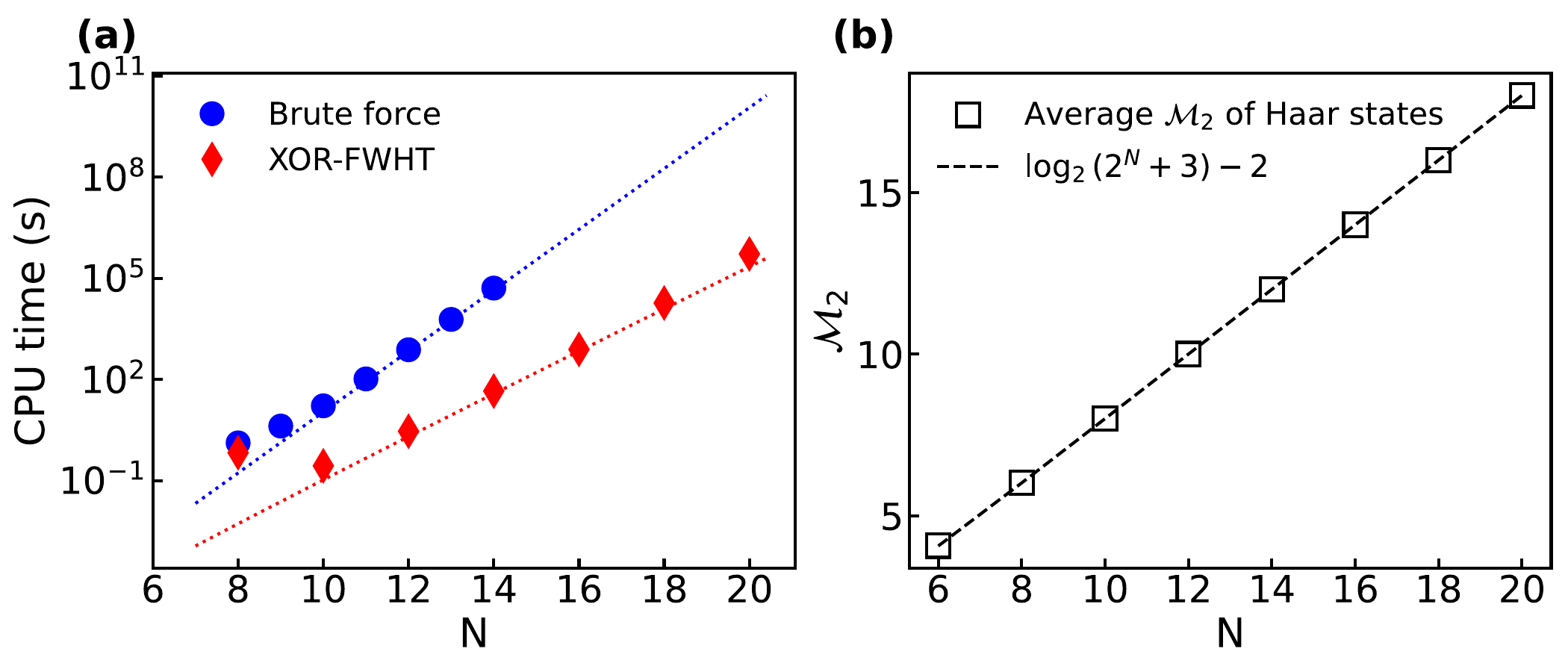}
\caption{(a) Exponential speedup of our XOR-FWHT algorithm (red) of the CPU time required for evaluating $\mathcal{M}_2$ of random pure states, in comparison with that from brute-force (blue). The dashed lines indicate the corresponding theoretical scaling behaviors, $\mathcal{O}(8^N)$ and $\mathcal{O}(N4^N)$, respectively. (b) The average 2-SRE $\mathcal{M}_2$ of Haar-random pure states as a function of the system size $N$. For each $N$, we independently sample $20-100$ Haar-random states and compute the average. The dashed line represents the theoretically predicted behavior, $\log_2(2^N+3)-2$, reported in Ref.~\cite{PhysRevB.111.054301}.}
\label{fig:1}
\end{figure}

\noindent{\textcolor{purple}{\textit{Algebraic Reformulation of the $2$-SRE.---}}} 
The SRE of pure states, originally introduced in Ref.~\cite{PhysRevLett.128.050402}, is defined from the full set of Pauli expectation values as
\begin{equation}
\mathcal M_\alpha (|\psi\rangle)
=
\frac{1}{1-\alpha}\log\!\left( \frac{1}{d}
\sum_{P \in \mathcal P_N}
\bigl| \langle \psi | P | \psi \rangle \bigr|^{2\alpha}
\right),
\label{eq:sre}
\end{equation}
where $N$ is the number of qubits, $d=2^N$, and $\alpha$ is the R\'enyi index. Here, $\mathcal P_N$ denotes the $N$-qubit Pauli group generated by tensor products of the single-qubit operators $\{\mathbf{1},X,Y,Z\}$. Since $|\mathcal P_N|=4^N$, a direct evaluation of Eq.~\eqref{eq:sre} is generically exponentially costly. For $\alpha \ge 2$, the SRE is a magic monotone~\cite{PhysRevLett.128.050402,Haug2023stabilizerentropies,PhysRevA.110.L040403}. In particular, it satisfies several key properties:
(i) \emph{faithfulness}, namely, $\mathcal M_\alpha(|\psi\rangle)=0$ if and only if $|\psi\rangle$ is a stabilizer state;
(ii) \emph{invariance under Clifford transformations}, i.e., for all $C\in\mathcal C_N$,
$\mathcal M_\alpha(C|\psi\rangle)=\mathcal M_\alpha(|\psi\rangle)$; and
(iii) \emph{additivity} under tensor products,
$
\mathcal M_\alpha(|\phi\rangle\otimes|\psi\rangle)
=
\mathcal M_\alpha(|\phi\rangle)+\mathcal M_\alpha(|\psi\rangle).
$
Here, $\mathcal C_N$ denotes the Clifford group. 
%These properties make the SRE a natural and physically well-motivated measure for characterizing magic in pure-state quantum many-body systems.

In this work, we focus on the case $\alpha=2$, which already corresponds to the simplest nontrivial and most widely studied member of the SRE family. Starting from the definition of the SRE in Eq.~\eqref{eq:sre}, we specialize to the case $\alpha=2$ and briefly derive the working expression used below, leaving the more detailed steps to the Supplemental Information. To expose the hidden convolution structure underlying the 2-SRE, we first rewrite the Pauli-string summation in a binary bitstring representation. Noting that $Y=iXZ$, and that the overall phase of a Pauli string does not affect $|\langle\psi|P|\psi\rangle|^4$, we may, without loss of generality, ignore the global phase and retain only the $X^xZ^z$ structure. Any $N$-qubit Pauli string can then be written uniquely as
% \begin{equation}
$
P = X_1^{x_1} \cdots X_N^{x_N}
    Z_1^{z_1} \cdots Z_N^{z_N}
\equiv X^x Z^z,
$
% \end{equation}
where
$
x:=(x_1,\dots,x_N),\;
z:=(z_1,\dots,z_N)
\in \mathbb{Z}_2^N
$.
Therefore, summing over all Pauli strings $P \in \mathcal P_N$
is equivalent to summing over all $x,z \in \mathbb Z_2^N$.  We write the quantum state in the computational bitstring basis as
$|\psi\rangle=
\sum_{t \in \mathbb{Z}_2^N}
\phi_t\, |t\rangle,$
where $|t\rangle$ denotes the computational basis state labeled by the length-$N$ bitstring $t=(t_1,\dots,t_N)$.
The actions of the Pauli operators $X^x$ and $Z^z$ on the computational basis are respectively given by
$
X^x |t\rangle = |t \oplus x\rangle, 
Z^z |t\rangle = (-1)^{z \cdot t}\, |t\rangle,
$
where $\oplus$ denotes bitwise addition modulo $2$,
and $z \cdot t\!=\!\sum_{j=1}^N z_j t_j \pmod 2$
is the binary inner product. Hence,
\begin{equation}
P|\psi\rangle
=
\sum_{t}
\phi_t\,(-1)^{z\cdot t}\,
|t\oplus x\rangle. \label{eq:wht}
\end{equation}
Taking the overlap with $\langle\psi|$, we obtain
\begin{equation}
\langle \psi | P | \psi \rangle
=
\sum_{t}
(-1)^{z \cdot t}\,
\phi_{t \oplus x}^*
\phi_t .
\label{eq:Pauli_expect}
\end{equation}
This expression already makes the key structure visible: for fixed $x$, the dependence on $z$ takes the form of a Walsh-Hadamard kernel, while the amplitudes appear in an XOR-shifted correlation $\phi_{t\oplus x}^*\phi_t$. The binary representation of Pauli strings and their action on computational-basis states are standard; the key new observation of this work is that, when inserted into the quartic 2-SRE, these ingredients can be systematically reorganized into XOR-correlation and convolution structures over $\mathbb Z_2^N$.

To see the first nontrivial consequence of this reformulation, let
\(
A_x(z):=\sum_t (-1)^{z\cdot t}\phi_{t\oplus x}^*\phi_t
\).
Then the 2-SRE involves sums of $|A_x(z)|^4$ over all $z\in\mathbb Z_2^N$. Expanding the fourth power and using the orthogonality of the Walsh-Hadamard characters,
\(
\sum_{z\in\mathbb Z_2^N}(-1)^{z\cdot s}=2^N\delta_{s,0},
\)
one finds
\begin{equation}
\sum_{z} |A_x(z)|^4
=
2^N
\sum_{\substack{
t_1,t_2,t_3,t_4\\
t_1\oplus t_2\oplus t_3\oplus t_4 = 0
}}
g_x(t_1)\, g_x^*(t_2)\,
g_x(t_3)\, g_x^*(t_4),
\label{eq:XOR_constraint0}
\end{equation}
where $g_x(t):=\phi_{t\oplus x}^*\phi_t$. Thus, the summation over $z$ projects the originally unconstrained quartic sum onto an XOR-constrained subspace. This XOR constraint is the crucial structural step: it is what ultimately allows the quartic Pauli-string summation to be recast into a form amenable to Walsh-Hadamard acceleration.

Proceeding from Eq.~\eqref{eq:XOR_constraint0}, one can further reorganize the constrained sum into XOR-correlation functions and then apply the convolution and Parseval properties of the Walsh-Hadamard transform. This yields the exact formula
\begin{equation}
\mathcal M_2(|\psi\rangle)
=
-\log\!\left(
\frac{1}{d}
\sum_{x, k}{|\text{FWHT}(\phi_{x\oplus k}^* \phi_x)|^4} \right),
\label{eq:final1}
\end{equation}
where $k\in \mathbb{Z}_2^N$, and FWHT~\cite{1671278} denotes the fast numerical implementation of the WHT. For an input of length $d=2^N$, direct evaluation of the WHT scales as $\mathcal O(d^2)$, whereas the FWHT reduces this to $\mathcal O(d\log d)=\mathcal O(N2^N)$ with modest additional memory overhead. A more detailed derivation is provided in Sec.~\ref{app:1} of the Supplemental Information. Equation~\eqref{eq:final1} is the key outcome of the derivation: it shows that the exact evaluation of the 2-SRE for a generic state vector can be reduced to repeated FWHTs acting on XOR-shifted amplitude correlators. 
%More generally, this derivation suggests a broader strategy for simplifying other Pauli-based quantities whenever analogous convolution structures can be identified.
\newline

\noindent{\textcolor{purple}{\textit{XOR-FWHT Algorithm for the $2$-SREs.---}}} By using Eq.~\eqref{eq:final1}, we design our Algorithm~\ref{alg1} for $\mathcal{M}_2$ of any arbitrarily chosen state.  
It is worth emphasizing that the bitstrings $x$ and $k$ are represented by their integer encodings in the numerical implementation, which is straightforward for a computer to handle.  
The implementation requires performing $d=2^N$ FWHTs, each of length $d$, leading to an overall time complexity of $\mathcal{O}(2^N \times N2^N)=\mathcal{O}(N4^N)$.  
The additional memory overhead originates from storing the reordered conjugate vector $\overline{\phi[\bm{x}\oplus k]} \quad (x=0,\ldots,d-1)$; this overhead does not accumulate with $k$ and is therefore acceptable.  
Moreover, the computations corresponding to different values of $k$ are mutually independent, making the algorithm naturally well-suited to parallel implementation.  
Implementation details are provided in Sec.~\ref{app:sec2} of the Supplemental Information.

\begin{algorithm}[H]
\caption{XOR-FWHT Algorithm of the  $\mathcal{M}_2$ }\label{alg1}
\begin{flushleft}
\textbf{Input}: normalized state vector $\phi[0],\ldots,\phi[d-1]$in the computational
basis , where $d=2^N$.\\
\textbf{Output}: $\mathcal{M}_2$.
\end{flushleft}
\begin{algorithmic}[1]
\State $r \gets 0$
\State $\bm{x} \gets [0,1,\ldots,d-1]$ 
\For{$k=0$ to $d-1$} 
    \State \textbf{construct} length-$d$ complex array $G$ by
    \[
      G[x] \gets \overline{\psi[x\oplus k]}\,\psi[x]\qquad (x=0,\ldots,d-1)
    \]
    \State $\hat{G} \gets \mathrm{FWHT}(G)$
   
    \State $r \gets r + \sum_{u=0}^{d-1} |\hat{G}[u]|^4$
\EndFor
\State \Return $\mathcal{M}_2 \gets -\log_2\!\bigl(r/d\bigr)$
\end{algorithmic}
\end{algorithm}

We benchmarked the performance of the algorithm on a workstation equipped with two Intel Xeon Platinum 8352V processors (36 cores per CPU, 72 cores in total). In the benchmarks, the input states were Haar-random states, and we measured the wall-clock time for a single evaluation of the 2-SRE.
For different system sizes $N$, the runtimes are as follows: approximately $0.88$ seconds for $N=14$; about $12$ seconds for $N=16$; about $300$ seconds for $N=18$; and about $2$ hours for $N=20$.
These numerical results are consistent with the theoretical time complexity $\mathcal{O}(N4^N)$ of the algorithm. In Fig.~\ref{fig:1}(a), we compare the CPU time of Algorithm~\ref{alg1} with that of a direct brute-force approach. Compared with the brute-force approach, our algorithm achieves an exponential speedup. The corresponding implementation is also provided in the Supplemental Information.

As a validation of the exactness of the algorithm, we evaluate the 2-SRE of Haar-random states. A Haar-random pure state is a pure quantum state drawn uniformly from Hilbert space according to the Haar measure. In many-body quantum physics, Haar-random pure states serve as benchmark models for characterizing the typical properties of generic pure states in quantum many-body systems~\cite{PhysRevLett.126.030601}. Haar-random states exhibit near-Page-law volume-law entanglement, which causes tensor-network-based methods for evaluating $\mathcal{M}_2$ to encounter an exponentially growing bond-dimension bottleneck as the system size increases.
By contrast, the XOR-FWHT algorithm proposed in this work does not rely on the bond dimension of a tensor-network representation, but instead acts directly on the state vector itself, thereby enabling the efficient evaluation of the magic measure $\mathcal{M}_2$ for Haar-random states. As shown in Fig.~\ref{fig:1}(b), the numerical results are in good agreement with the theoretical prediction~\cite{PhysRevB.111.054301}, thereby validating the applicability of the algorithm to highly entangled states.\\

% \\

%\noindent{\textcolor{purple}{\textit{Applications in Physical Examples.---}}
\noindent{\textcolor{purple}{\textit{Physical Application Examples.---}}
In this section, we numerically present several representative physical examples by using algorithm~1 for evaluating 2-SRE.

\begin{figure}[bt]
\includegraphics[width=\linewidth]{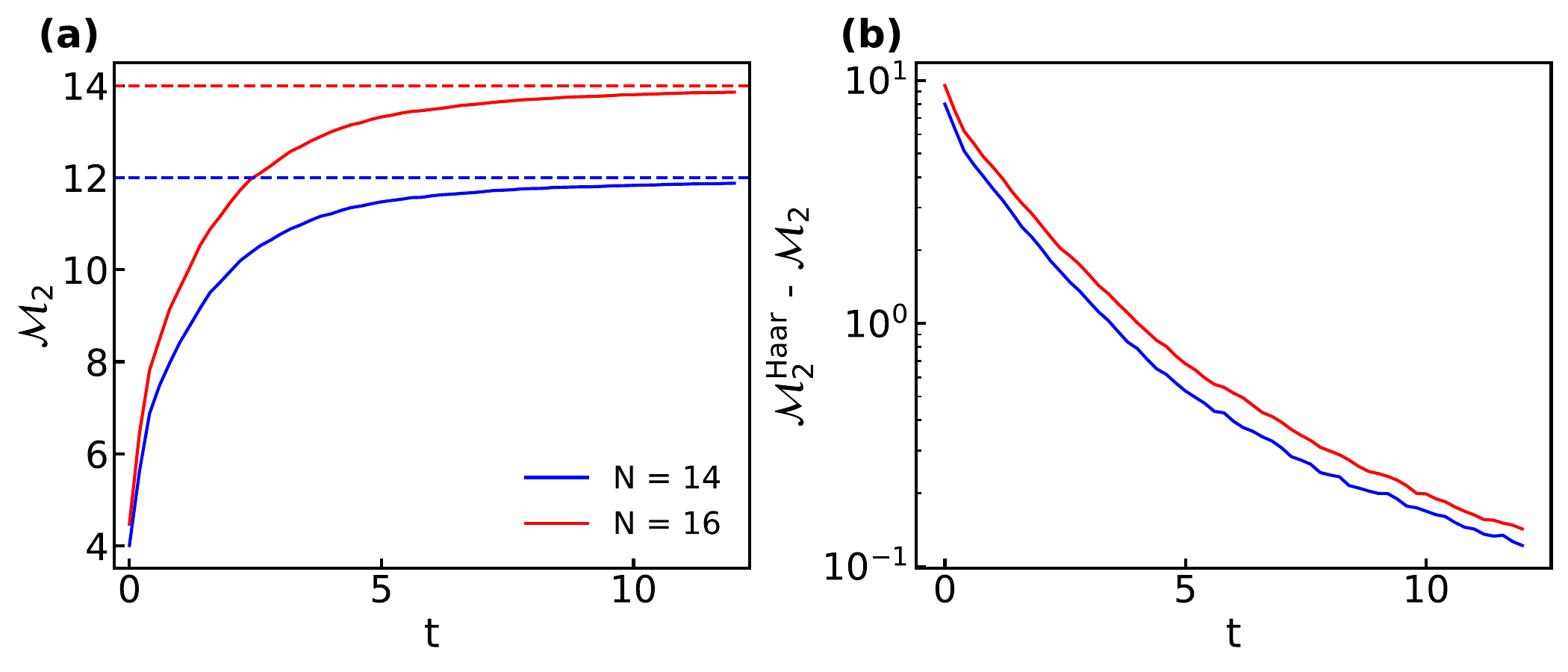}
\caption{ Time evolution of the 2-SRE $\mathcal{M}_2$ for $N\!=\!14$ and $16$ spins evolved under the TFIM with a longitudinal field, computed using our XOR-FWHT algorithm with time step $\Delta t = 0.2$. 
(a) Starting from $100$ random product states, $\mathcal{M}_2(t)$ rapidly grows and saturates within the accessible time window at values slightly below the Haar-random benchmark (dashed line). 
(b) The long-time behavior of $\mathcal{M}_2$ is resolved with sufficient accuracy to clearly show that it remains consistently below $\mathcal{M}_2^\text{Haar}$ throughout the time range considered.}
\label{fig:2}
\end{figure}

\noindent{{\textit{Example 1.~Hamiltonian quench dynamics. --}} We consider the transverse-field Ising model with an additional longitudinal field~\cite{imbrie2016many,PhysRevA.3.786,de2024absence,10.21468/SciPostPhysLectNotes.82,LIEB1961407,y9r6-dx7p,xfp5-hhs4} (TFIM with longitudinal field), whose Hamiltonian is given by
\begin{equation}
H_{\mathrm{TFIM+LF}}
=
- J \sum_{i=0}^{N-1} Z_i Z_{i+1}
- h_x \sum_{i=0}^{N-1} X_i
- h_z \sum_{i=0}^{N-1} Z_i ,
\label{eq:TFIM_LF_OBC}
\end{equation}
where we set $J=1$, $h_x=1.5$, and $h_z=1.5$, and consider system sizes $N=14$ and $16$.

The initial state is chosen as a tensor product of random single-qubit states,
\begin{equation}
\ket{\psi_0}
=
\bigotimes_{i=1}^{N}
\left(
\cos\frac{\theta_i}{2}\,\ket{\uparrow}
+
\mathrm{e}^{\mathrm{i}\phi_i}
\sin\frac{\theta_i}{2}\,\ket{\downarrow}
\right),
\label{eq:random_product_state}
\end{equation}
where $\theta_i \in [0,\pi]$ and $\phi_i \in [0,2\pi)$ are independent random variables. Although these initial states are unentangled, they are generally non-stabilizer states and therefore carry local non-stabilizer resources. Starting from $100$ different random initial states, we perform the real-time evolution using a matrix-exponential-action method inspired by the Krylov-subspace approach. In practice, we implement $e^{-iHt}|\psi_0\rangle$ via \texttt{scipy.sparse.linalg.expm\_multiply}. We then compute $\mathcal{M}_2$ during the evolution and average the results over all initial states.

The resulting dynamical behavior is shown in Fig.~\ref{fig:2}. We observe that, within the system sizes and time window considered, $\mathcal{M}_2(t)$ grows rapidly with time and gradually approaches saturation, while remaining slightly below the corresponding Haar-random-state benchmark throughout the evolution [see Fig.~\ref{fig:2}(a)]. The asymptotic behavior is shown more clearly in Fig.~\ref{fig:2}(b). Compared with the approximate sampling approaches previously used in the literature, such a subtle deviation from the Haar value is difficult to resolve; at that stage, approximate sampling was necessary due to the absence of an exact method with exponential speedup~\cite{y9r6-dx7p}. We conjecture that this is related to the intrinsic statistical errors of numerical methods based on finite sampling, which make such small deviations in $\mathcal{M}_2$ difficult to detect. By contrast, the exact XOR-FWHT algorithm employed here enables high-precision evaluation of $\mathcal{M}_2$, thereby allowing these subtle but physically meaningful deviations to be clearly resolved. This provides a new numerical tool for further investigating the fine structure of magic resources in dynamical processes.\\

% \subsection{Example 2.~random unitary circuit quench dynamics}

%
\begin{figure}[bt]
\includegraphics[width=\linewidth]{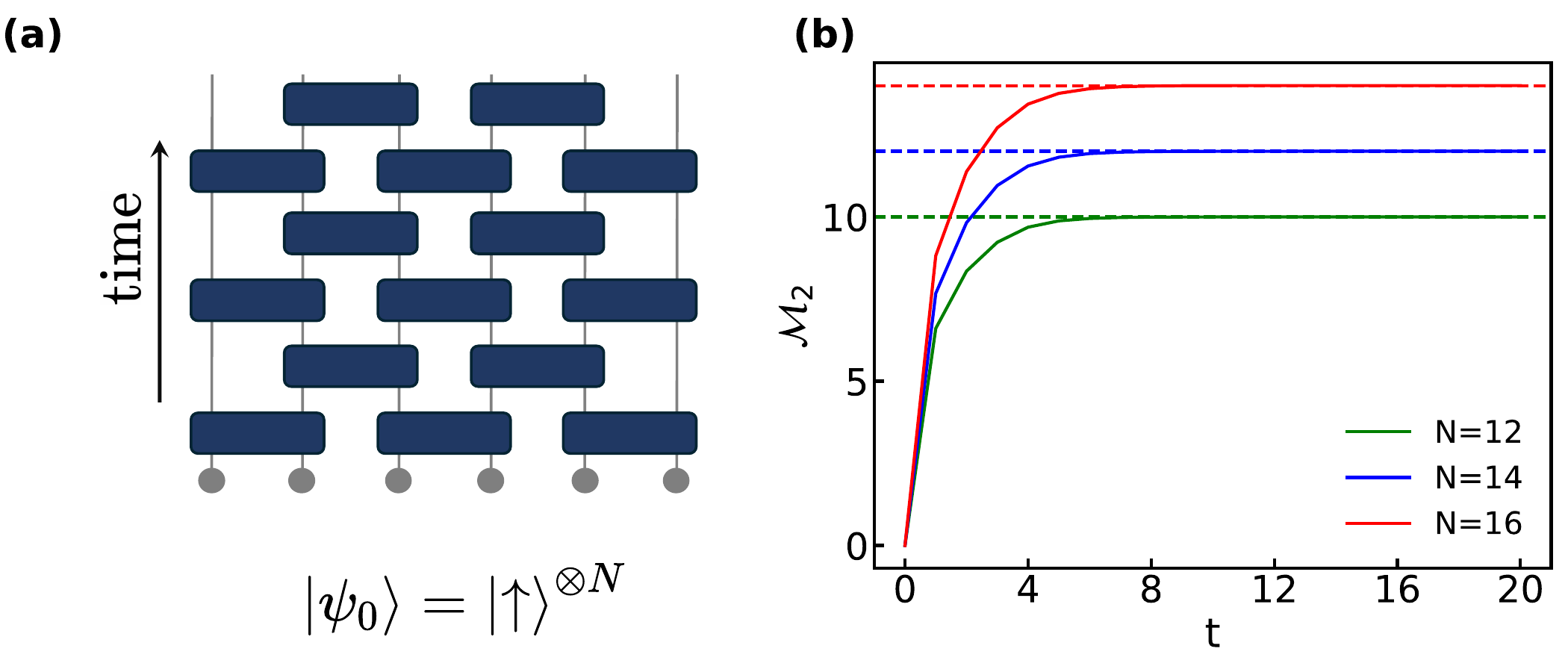}
\caption{(a) Schematic of the 1D brickwork random unitary circuit. The system is initialized in the product state $|\psi_0\rangle = |\uparrow\rangle^{\otimes N}$ and evolves under layers of nearest-neighbor Haar-random two-qubit gates. (b) Dynamics of the 2-SRE $\mathcal{M}_2$ for system sizes $N=12,14,16$, obtained by first simulating the circuit evolution and then feeding the resulting state vectors into our XOR-FWHT algorithm, with averaging over 100 circuit realizations. The dashed horizontal lines indicate the corresponding Haar-random-state benchmarks. }
\label{fig:3}
\end{figure}

\noindent{{\textit{Example 2.~random unitary circuit quench dynamics --}} We extend our analysis to dynamical quantum states generated by local random unitary circuits. Unlike Haar-random states sampled directly from Hilbert space, these states are produced progressively from a simple product state through nonequilibrium unitary evolution. As such, they exhibit generic features of many-body dynamics. Random unitary circuits themselves constitute a standard framework for the study of nonequilibrium many-body physics, and have been widely employed to characterize operator spreading and scrambling~\cite{Nahum2018,Swingle2018NatPhys,SekinoSusskind2008JHEP}, entanglement growth, equilibration, and typicality~\cite{PhysRevLett.126.030601,Fisher2023,hunterjones2019unitarydesignsstatisticalmechanics}, as well as the emergence of approximate unitary designs and Haar-like behavior~\cite{HarrowLow2009CMP,BrandaoHarrowHorodecki2016CMP,Haferkamp2022randomquantum}. More recently, they have also become an important platform for investigating the dynamics of magic~\cite{Turkeshi_2024_2,szombathy2025independentstabilizerrenyientropy,maity2025localspreadingstabilizerrenyi}. Therefore, studying the time evolution of the 2-SRE in this setting enables us to track how the 2-SRE gradually approaches its Haar-typical value during the dynamical process. Specifically, we consider dynamical quantum states generated by a one-dimensional nearest-neighbor brickwork random unitary circuit on an $N$-qubit chain with open boundary conditions, initialized in the product state $|\psi_0\rangle=\ket{\uparrow}^{\otimes N}$. At each discrete time step, the evolution consists of two successive layers of two-qubit unitary gates acting on the even and odd bonds, respectively; all two-qubit gates are sampled independently from the Haar measure on $U(4)$. The quantum state then evolves according to
$
|\psi_t\rangle = U_t |\psi_{t-1}\rangle,
$
where $U_t$ denotes the brickwork random circuit for the $t$-th time step. At each time step, we record the full quantum state and evaluate the corresponding 2-SRE.

As illustrated in Fig.~\ref{fig:3}(b), we first simulate the random-circuit dynamics for system sizes $N=12,14,16$ using \texttt{TensorCircuit-NG}~\cite{Zhang_2023,zhang2026tensorcircuit}, and then feed the resulting state vectors at each discrete time step into our XOR-FWHT algorithm to evaluate $\mathcal{M}_2$ exactly. As the circuit evolves, the state rapidly develops strong entanglement and gradually approaches Haar-like behavior, making the evaluation of $\mathcal{M}_2$ increasingly challenging for methods that rely on low-entanglement tensor-network representations. For all system sizes considered, $\mathcal{M}_2$ grows rapidly from the initial product state and saturates near the corresponding Haar-random-state benchmark at long times. This example extends the Haar-state benchmark to a genuinely dynamical setting and shows that our method can efficiently and accurately evaluate the 2-SRE for generic random-circuit states while directly resolving their approach to Haar-typical magic.
\\

\noindent{\textcolor{purple}{\textit{Conclusion and Outlook.---}} In this work, we have proposed a fast and exact algorithm for evaluating medium-scale 2-SRE $\mathcal{M}_2$, with computational complexity $\mathcal{O}(N4^N)$, thereby achieving an exponential speedup over the brute-force cost $\mathcal{O}(8^N)$. Starting from the definition of SRE, we reformulated it exactly into a deterministic form based on XOR convolution and the Walsh-Hadamard transform, without introducing any approximation, stochastic sampling, or additional assumptions on the structure of the quantum state, and this formulation can be implemented efficiently via the FWHT. Numerical results show that the algorithm not only accurately reproduces the theoretical behavior of $\mathcal{M}_2$ for Haar-random states, but also resolves subtle deviations that are difficult to capture clearly with sampling-based approaches, while remaining applicable to the growth and saturation of $\mathcal{M}_2$ in more general dynamical settings. Taken together, these results establish the XOR-FWHT algorithm as a fast, exact, and broadly applicable numerical tool for characterizing non-stabilizer resources in generic quantum states and their dynamics, and provide a basis for further investigations into finer structures of magic dynamics.}

Looking ahead, several promising directions remain for further exploration.
One direct application of the present algorithm is the systematic study of the dynamical behavior of $\mathcal{M}_2$ in a variety of interacting quantum many-body models~\cite{PaviglianitiLamiColluraSilva2025PRXQ, CH4,njgn-fksh, xfp5-hhs4, CH5,1jzy-sk9r},
as well as its interplay with other physical quantities such as entanglement~\cite{PhysRevB.110.045101, viscardi2025interplayentanglementstructuresstabilizer,jplh-zl35} and operator spreading~\cite{dowling2025magic, Nahum2018, vonKeyserlingkRakovszkyPollmannSondhi2018PRX}.
From a computational perspective, combining the current framework with tensor-network representations~\cite{PhysRevB.107.035148, PhysRevLett.131.180401, PhysRevLett.133.010601, liu2025stabilizerrenyientropytranslationinvariant} is particularly appealing,
as this may further reduce the effective computational cost for large systems or weakly entangled states.
More broadly, we expect that the algorithmic ideas developed in this work can serve as building blocks
for the efficient evaluation of higher-order stabilizer R\'enyi entropies and related resource measures.\\

\noindent{\textit{Acknowledgements --}} J.-X.\ Zhong was supported by the National Natural Science Foundation of China (Grant Nos.\ 12374046 and 11874316), the Shanghai Science and Technology Innovation Action Plan (Grant No.\ 24LZ1400800), the National Basic Research Program of China (Grant No.\ 2015CB921103), and the Program for Changjiang Scholars and Innovative Research Teams in Universities (Grant No.\ IRT13093). C. H. Lee acknowledges support from the Ministry of Education, Singapore (MOE Tier-II Award No. MOE-T2EP50224-0007, WBS no. A-8003505-00-00). H.-Z.\ Li is supported by the China Scholarship Council (CSC) Scholarship (Grant No.\ 202506890103).\\

\paragraph{Note added:} After the first version of this manuscript was preprinted on arXiv (arXiv:2512.24685v1), two related studies, also posted later on arXiv as arXiv:2601.00761v1 and arXiv:2601.07824v1, appeared independently~\cite{xiao2026exponentially, sierant2026computing}.
% \vspace{0.1 cm}
\paragraph{Data Availability:} The numerical data that support the findings of this study are publicly available at \url{https://github.com/hhhhhh258/2sre}.

\bigskip
\bibliography{refs}

\clearpage
\onecolumngrid

\clearpage
\onecolumngrid
\begin{center}
\textbf{\large Supplemental information for:\\ ``A fast and exact approach for stabilizer R\'enyi entropy via the XOR–FWHT algorithm''}
\end{center}
\setcounter{equation}{0}
\setcounter{figure}{0}
\renewcommand{\theequation}{S\arabic{equation}}
\renewcommand{\thefigure}{S\arabic{figure}}

\begin{center}
Xuyang Huang,$^{1, *}$
Han-Ze Li,$^{1,2, *}$
Ching Hua Lee,$^{2, \dagger}$
and Jian-Xin Zhong$^{1, \ddagger}$
\vspace{0.3cm}

\textit{$^{1}$Institute for Quantum Science and Technology, Shanghai University, Shanghai 200444, China}\\
\textit{$^{2}$Department of Physics, National University of Singapore, Singapore 117542}\\

\end{center}
\vspace{0.5cm}

\section{Simplifying the Expression of the 2-SRE with the Walsh--Hadamard Transform (WHT)}\label{app:1}

We start from Eq.~\ref{eq:sre} and specialize to the case $\alpha=2$. The relevant properties of the WHT used below are summarized in Sec.~\ref{app:WHT_properties}. To keep this section self-contained, we briefly recall the notation introduced in the main text. After disregarding the overall phase of a Pauli string, any \(N\)-qubit Pauli string can be written as \(P=X^x Z^z\), where \(x,z\in\mathbb Z_2^N\). The quantum state is expanded in the computational basis as \(|\psi\rangle=\sum_{t\in\mathbb Z_2^N}\phi_t |t\rangle\), where \(\phi_t\) denotes the wavefunction amplitude and \(t\in\mathbb Z_2^N\) is a length-\(N\) binary bitstring. Here \(\oplus\) denotes bitwise addition modulo \(2\), and \(z\cdot t:=\sum_{j=1}^N z_j t_j \pmod 2\) denotes the binary inner product. 

Using \(X^x|t\rangle=|t\oplus x\rangle\) and \(Z^z|t\rangle=(-1)^{z\cdot t}|t\rangle\), one obtains
\(P|t\rangle=(-1)^{z\cdot t}|t\oplus x\rangle\), and hence
\(P|\psi\rangle=\sum_t \phi_t (-1)^{z\cdot t}|t\oplus x\rangle\).
Taking the overlap with \(\langle\psi|=\sum_u \phi_u^* \langle u|\), and using \(\langle u|t\oplus x\rangle=\delta_{u,t\oplus x}\), we arrive at the expectation value in the Pauli basis,
\begin{equation}
\langle \psi | P | \psi \rangle
=
\sum_{t}
(-1)^{z \cdot t}\,
\phi_{t \oplus x}^*
\phi_t .
\label{eq:Pauli_expect}
\end{equation}
Here \(x\) is precisely the binary exponent associated with the \(X\)-part of the Pauli string \(P=X^x Z^z\).

This expression exhibits an XOR-correlation structure,
namely a weighted sum over correlations between the wavefunction amplitudes \(\phi_t\)
and their bit-shifted counterparts \(\phi_{t \oplus x}\).
In the following derivation, we first fix \(x\) and perform the summation over \(z\).

\emph{Fourth-order modulus expansion and Walsh-Hadamard orthogonality.}
Define
$g_x(t) := \phi_{t \oplus x}^* \phi_t ,$
and introduce
$
A(z)
:= \sum_{t \in \mathbb{Z}_2^N}
(-1)^{z \cdot t}\, g_x(t),
$
so that
\begin{equation}
|\langle \psi | P | \psi \rangle|^4=|A(z)|^4
=
A(z)A^*(z)A(z)A^*(z).
\end{equation}
Expanding this expression explicitly yields
$
|A(z)|^4 =
\sum_{t_1,t_2,t_3,t_4}
(-1)^{z \cdot (t_1 \oplus t_2 \oplus t_3 \oplus t_4)}
\nonumber \times
g_x(t_1)\, g_x^*(t_2)\,
g_x(t_3)\, g_x^*(t_4).
$

Since the definition of the SRE involves a sum over all
\(z \in \mathbb Z_2^N\),
we can exploit the orthogonality relation of the Walsh-Hadamard transform
defined on the group \(\mathbb Z_2^N\),
\begin{equation}
\sum_{z \in \mathbb{Z}_2^N}
(-1)^{z \cdot s}
=
\begin{cases}
2^N, & s = 0, \\
0, & s \neq 0,
\end{cases}
\end{equation}
to explicitly carry out the summation over \(z\), obtaining
\begin{align}
\sum_{z} |A(z)|^4
&=
2^N
\sum_{\substack{
t_1,t_2,t_3,t_4\\
t_1\oplus t_2\oplus t_3\oplus t_4 = 0
}}
g_x(t_1)\, g_x^*(t_2)\,
g_x(t_3)\, g_x^*(t_4).
\label{eq:XOR_constraint}
\end{align}
The orthogonality of the Walsh-Hadamard transform projects the originally unconstrained
fourfold sum onto the subspace satisfying
\(t_1\oplus t_2\oplus t_3\oplus t_4 = 0\).
This XOR constraint is the key to the subsequent efficient numerical implementation.

\emph{XOR-convolution structure and dimensionality reduction.}
The constraint
\(t_1\oplus t_2\oplus t_3\oplus t_4 = 0\)
allows one to eliminate one variable, for instance, by writing
\(t_4 = t_1\oplus t_2\oplus t_3\).
Eq.~\eqref{eq:XOR_constraint} can thus be rewritten as
\begin{align}
\sum_{z} |A(z)|^4
&=
2^N
\sum_{t_1,t_2,t_3}
g_x(t_1)\, g_x^*(t_2)\,
g_x(t_3)\, g_x^*(t_1\oplus t_2\oplus t_3).
\label{eq:XOR_constraint1}
\end{align}
To make the XOR-convolution structure explicit, we introduce the new variable
\(s=t_2\oplus t_3\), which defines a one-to-one change of variables since
\(t_3=t_2\oplus s\). Using the elementary XOR identities
\(t_2\oplus t_2=0\) and commutativity, one finds
\(
t_1\oplus t_2\oplus t_3
=
t_1\oplus s
\).
Therefore,
\begin{align}
\sum_{z} |A(z)|^4
&=
2^N
\sum_{t_1,t_2,s}
g_x(t_1)\, g_x^*(t_2)\,
g_x(t_2\oplus s)\, g_x^*(t_1\oplus s)
\nonumber\\
&=
2^N
\sum_{s\in\mathbb Z_2^N}
\left(
\sum_{t_1\in\mathbb Z_2^N}
g_x(t_1)\,g_x^*(t_1\oplus s)
\right)
\left(
\sum_{t_2\in\mathbb Z_2^N}
g_x^*(t_2)\,g_x(t_2\oplus s)
\right)
\nonumber\\
&=
2^N
\sum_{s\in\mathbb Z_2^N}
\left|
\sum_{t\in\mathbb Z_2^N}
g_x(t)\,g_x^*(t\oplus s)
\right|^2 .
\label{eq:final_convolution_pre}
\end{align}
This naturally motivates the definition
$
C_x(s)
:=
\sum_{t \in \mathbb{Z}_2^N}
g_x(t)\, g_x^*(t \oplus s),
$
where \(s \in \mathbb{Z}_2^N\), so that Eq.~\eqref{eq:final_convolution_pre} becomes
\begin{align}
\sum_{z} |A(z)|^4
&=
2^N
\sum_{s \in \mathbb{Z}_2^N}
|C_x(s)|^2 .
\label{eq:final_convolution}
\end{align}

\emph{Exploiting properties of the WHT.}
We apply the WHT to \(C_x(s)\), and define \(G_x := \text{WHT}(g_x)\).
Using property~\ref{app:conj}, we obtain
$
\text{WHT}(C_x) =  \sqrt{2^N} G_x \cdot G_x^* = \sqrt{2^N} |G_x|^2 .
$
Applying the Parseval identity of the WHT~\ref{app:par} to Eq.~\eqref{eq:final_convolution} yields
$
\sum_s |C_x(s)|^2 =  \sum_k |\text{WHT}(C_x)(k)|^2 .
$
This further leads to
$
\sum_s |C_x(s)|^2 = {2^N} \sum_k |G_x(k)|^4 .
$
Then we obtain (there is also a factor of $2^N$ in Eq.~\eqref{eq:final_convolution})
\begin{equation}
\sum_{P \in \mathcal P_L}
\bigl| \langle \psi | P | \psi \rangle \bigr|^4 = 4^N
\sum_{x, k}{|\text{WHT}(\phi_{x\oplus k}^* \phi_x)|^4} .
\label{eq:WHT_final}
\end{equation}
We note that the FWHT provides an efficient numerical implementation of the WHT. While a direct evaluation of the WHT for a vector of length $d=2^N$ requires $O(d^2)$ operations, the FWHT reduces this cost to $\mathcal{O}(d\log_2 d)=\mathcal{O}(N2^N)$ by exploiting the recursive structure of the Hadamard kernel. This reduction is essential for the practical implementation of the present algorithm. More detailed algorithmic remarks on the FWHT are given in Sec.~\ref{app:fwht}. By recasting Eq.~\eqref{eq:WHT_final}, one obtains the final expression:

\begin{equation}
\sum_{P \in \mathcal P_L}
\bigl| \langle \psi | P | \psi \rangle \bigr|^4 = 
\sum_{x, k}{|\text{FWHT}(\phi_{x\oplus k}^* \phi_x)|^4} .
\label{app: finial0}
\end{equation}

In this derivation, we identify the hidden  XOR-convolution structures underlying the quartic Pauli-string summation of the 2-SRE, and systematically reorganize them into a form amenable to Walsh-Hadamard acceleration. This step allows the analytic structure of the 2-SRE to be directly translated into a fast and exact FWHT algorithm applicable to generic many-qubit states.

\section{Properties of the Walsh-Hadamard transform (WHT)}
%CH: it is just a summary of its properties, don't think it qualifies as an intro 

\label{app:WHT_properties}
To facilitate the reader’s understanding of the logical structure of the derivations presented , we summarize in this section the properties and features of the WHT used in this work.

\subsection{Definition}
\noindent We first define the WHT of a function
$f : \{0,1\}^N \to \mathbb{C}$ as
\begin{equation}
\hat f(k)
=
\frac{1}{\sqrt{2^N}}\sum_{x \in \{0,1\}^N}
(-1)^{k \cdot x} f(x),
\end{equation}
where $k \cdot x$ denotes the mod-$2$ inner product.

\subsection{Property of Involution and Inverse Transform}\label{app:inv}
\noindent The WHT satisfies
\begin{equation}
\label{eq:WHT_involution}
\mathrm{WHT}\bigl(\mathrm{WHT}(f)\bigr)
=
f.
\end{equation}
Therefore, the inverse transform is given by
\begin{equation}
\label{eq:IWHT}
\mathrm{IWHT}(f)
=
\mathrm{WHT}(f).
\end{equation}

\subsection{Properties of the XOR-Convolution Theorem}\label{app:XORc}

\noindent We define the XOR convolution as
\begin{equation}
\label{eq:XOR_conv}
(f *_\oplus g)(x)
=
\sum_{y\in\mathbb Z_2^N} f(y)\, g(x\oplus y),
\end{equation}
then
\begin{equation}
\label{eq:WHT_conv}
\mathrm{WHT}(f *_\oplus g)
=
\sqrt{2^N} \mathrm{WHT}(f)\cdot  \mathrm{WHT}(g),
\end{equation}
where the symbol $\cdot$ on the right-hand side denotes pointwise multiplication of vectors.

\subsubsection{Property I: Compatibility with Complex Conjugation}\label{app:conj}

\noindent For any complex-valued vector \(f\), one has
\begin{equation}
\label{eq:WHT_conj}
\mathrm{WHT}(f^*)(k)
=
\bigl(\mathrm{WHT}(f)(k)\bigr)^*.
\end{equation}
Consequently,
\(\mathrm{WHT}(f)\,\mathrm{WHT}(f^*)=|\mathrm{WHT}(f)|^2\).

\subsubsection{Property II: Parseval’s Theorem}\label{app:par}

\noindent The WHT preserves the \(\ell_2\) norm:
\begin{equation}
\label{eq:Parseval_WHT}
\sum_{x\in\mathbb Z_2^N} |f(x)|^2
=
\sum_{k\in \mathbb Z_2^N}
\bigl|\mathrm{WHT}(f)(k)\bigr|^2.
\end{equation}

% \subsection{ Numerical Algorithm for the fast WHT (FWHT) }\label{app:fwht}
% The WHT admits an efficient numerical
% implementation known as the FWHT.
% Rather than defining a new transform, the FWHT computes the WHT by
% exploiting the recursive structure of the Hadamard matrix.
% In our implementation, the FWHT evaluates the unnormalized transform,
% which is related to the normalized WHT by
% \begin{equation}
% \mathrm{FWHT}(f)=\sqrt{2^N}\,\mathrm{WHT}(f).
% \end{equation}

% Concretely, the FWHT proceeds via iterative butterfly updates of the form \CH{CH: Explain more how/why that work}
% \begin{equation}
% (a, b) \mapsto (a+b, a-b),
% \end{equation}
% applied recursively over $N$ levels for functions defined on
% $\{0,1\}^N$, resulting in an overall computational complexity of
% $O(N2^N)$.
% Implementation details are provided in Appendix~\ref{app:sec2}.

\subsection{Numerical Algorithm for the fast WHT (FWHT)}
\label{app:fwht}

The WHT admits an efficient numerical implementation known as the FWHT. The FWHT evaluates the same Walsh--Hadamard transform by exploiting the recursive block structure of the Hadamard matrix. In our implementation, the FWHT computes the unnormalized transform, which is related to the normalized WHT by
\begin{equation}
\mathrm{FWHT}(f)=\sqrt{2^N}\,\mathrm{WHT}(f).
\end{equation}

The efficiency of the FWHT follows from the recursive identity
\begin{equation}
H_{2^N}
=
\begin{pmatrix}
H_{2^{N-1}} & H_{2^{N-1}}\\
H_{2^{N-1}} & -H_{2^{N-1}}
\end{pmatrix},
\end{equation}
where \(H_{2^N}=H_2^{\otimes N}\) denotes the unnormalized Hadamard matrix of size \(2^N\), with
\(
H_2=
\begin{pmatrix}
1 & 1\\
1 & -1
\end{pmatrix}.
\)
This block decomposition shows that the transform of a length-\(2^N\) vector can be built recursively from transforms of its two length-\(2^{N-1}\) halves, followed by pairwise sum-and-difference combinations. At the level of vector entries, this gives the elementary butterfly update
$
(a,b)\mapsto (a+b,a-b).
$

For functions defined on \(\{0,1\}^N\), these butterfly operations are applied hierarchically over \(N\) stages, each stage corresponding to one binary index. At every stage, all \(2^N\) entries are updated once by constant-cost additions and subtractions, so the cost per stage is \(\mathcal{O}(2^N)\). Repeating this procedure over all \(N\) stages yields the total computational complexity $\mathcal{O}(N2^N)$.

This reduction from the direct \(\mathcal{O}(4^N)\) evaluation is essential for the practical implementation of the present algorithm. Implementation details are provided in Sec.~\ref{app:sec2}.

\subsection{ Python Implementation }\label{app:sec2}
% \CH{CH: actually, no need for a new section just to have the code}

% In this appendix, we present the XOR-FWHT algorithm used in this work, together with the brute-force approach.

\textbf{XOR-FWHT approach}

\begin{lstlisting}[language=Python]
import numpy as np

def FWHT(a):
    n = a.shape[0]
    h = 1
    while h < n:
        for i in range(0, n, 2*h):
            u = a[i:i+h]
            v = a[i+h:i+2*h]
            u[:], v[:] = u+v, u-v
        h *= 2

def xor_fwht(psi):
    d = psi.shape[0]
    x = np.arange(d, dtype=np.int64)
    total = 0.0
    psi_c = np.conjugate(psi)
    for k in range(d):
        A = psi_c[x^k] * psi    
        FWHT(A)                           
        total += (np.sum(np.abs(A) ** 4))
    return -np.log2(total/d)

\end{lstlisting}
\vspace{.5 cm}
\textbf{Brute-force apprach}
\begin{lstlisting}[language=Python]
import numpy as np

def dot(z, t, N):
    n = z & t
    res = 0
    for i in range(N):
        res += ((n>>i) & 1)
    return res

def brute_force(psi, N):
	d = psi.size
	psi_c = np.conjugate(psi) 
	sum_ = 0.0
	for x in range(d):
		for z in range(d):
			val = 0.0
			for t in range(d):
				val += (-1)**dot(z, t, N) * psi_c[t^x] * psi[t]
			sum_ += np.abs(val)**4
	return -np.log2(sum_/d)

\end{lstlisting}

For clarity, the code snippets shown below present only the core computational logic at the Python/NumPy level. In the actual implementation, we further employ Numba just-in-time (JIT) compilation and parallelization to accelerate the computation.

% \noindent It should be emphasized that, in the practical implementation of this code, we also employ JIT \CH{CH: whats JIT} and parallelization techniques, enabling us to fully utilize multicore CPU resources.

\end{document}